# Towards perfection: Kapellasite, $Cu_3Zn(OH)_6Cl_2$, a new model S=1/2 kagome antiferromagnet

R H Colman, C. Ritter, A.S. Wills

*Department of Chemistry, UCL, Christopher Ingold Laboratories, 20 Gordon Street, London, UK, WC1H 0AJ*

*The Institut Laue-Langevin, 6 Rue Jules Horowitz, BP156, 38042 Grenoble, Cedex 9, France*

While much is becoming known experimentally about the effects of frustration in the classical limit, such as in the 3-dimensional pyrochlores[1,2,3] and 2-dimensional (2D) kagome lattice[4,5,6,7,8], what happens in the S=1/2 quantum limit is little characterized due to the scarcity of model materials. This has led to one of the most enduring problems in condensed matter physics: the search for the 'resonating valence bond' (RVB) state. The state was proposed by Anderson over two decades ago to underpin the transition to BCS superconductivity.[9] The RVB is a quantum state made by the spin pairing of electrons on different sites. Its highly entangled nature makes it an essential reference for the physical sciences, and for the current efforts in the development of states that can be exploited for quantum computing[10]

The $S$=1/2 kagome antiferromagnet (KAFM) is a good candidate for the realization of the RVB state[11,12,13] and much excitement has been generated since the recent proposal of Herbertsmithite (also known as Zn–paratacamite) as a model $S$=1/2 KAFM.[14] The material, named after the mineralogist Herbert Smith, is a doped member of the atacamite familty in which one quarter of the sites of a pyrochlore-like lattice are occupied by diamagnetic $Zn^{2+}$: the remaining sites are occupied by the $Cu^{2+}$ spins to form the kagome net. While it was initially hailed as a 'perfect' KAFM, NMR studies have shown that there is appreciable disorder in the Cu/Zn partitioning[15] and that the Dzyaloshinski-Moriya (DM) interactions allowed in the kagome geometry play an important role.[16,17] Any intersite Cu/Zn substitution in Herbertsmithite is particularly significant as it may introduce coupling between the kagome planes that destroys the 2-dimensionality of the magnetic lattice. The local structure would then resemble the $Cu_4$ end-member clinoatacamite, γ-$Cu_2(OH)_3Cl$.[18,19,20,21]

In this paper we report the synthesis and preliminary magnetic characterization of the newly discovered mineral Kapellasite, a metastable polymorph of Herbertsmithite which shares its chemical formula, $Cu_3Zn(OH)_6Cl_2$.[22] Kapellasite is synthesized from $ZnCl_2$ solution and copper metal.[23] The crystal structure is a much better approximation to the KAFM with the 2D layers being only weakly coupled by O—H—Cl hydrogen bonds, giving rise to well defined cleavage planes along (001). The diamagnetic $Zn^{2+}$ acts to dilute what would otherwise be a triangular net of $Cu^{2+}$ to form a well defined kagome lattice (Figure 1). Whereas

Cu/Zn disorder in Herbertsmithite introduces local exchange between the kagome planes, the perturbative effect in Kapellasite is expected to be much smaller due to the quasi-2D nature of the structure. Further, the exotic properties relating to the RVB state of the KAFM are expected to survive doping: the doped KAFM is a model for the $CuO_2$ planes in the underdoped cuprate superconductors.[24] This quantum behavior under the influence of disorder is quite different to that seen on the classical KAFMs where disorder can lead to Néel order.[25,26]

Powder neutron diffraction (PND) taken with the D1a diffractometer at the ILL from a 0.3g sample of protonated Kapellasite showed that under these synthetic conditions only a single copper hydroxychloride phase is formed, and the presence of unreacted copper particles was not observable. Rietveld refinement of the PND data using FullProf[27] showed the structure to be in good agreement with that determined previously (Figure 2).[22] The crystal structure is expressed in space group P-3m1 and the refined crystallographic parameters are given in Supplementary Table 1. There was no evidence of an intersite substitution, either of the Zn/Cu or Cl/OH within that quality of our data and all occupancies were left at their ideal values. A high degree of segregation is expected from the very different local geometries of the two metal sites. As in Herbertsmithite, the Jahn-Teller distorted $Cu^{2+}$ resides more favorably on the tetragonally elongated 3$f$ site; the Zn is effectively segregated at the 1$b$ position and coordinated by a trigonally compressed octahedron of 6 equidistant Zn—O bonds. Bond Valence calculations using VaList[28] are in good agreement with expected values: the bond valence sums and bond angles are shown in Supplementary Tables 2 and 3, respectively.

The zero-field cooled (ZFC) and field-cooled (FC) magnetic susceptibility measured in fields between 50 and 70 000~G do not indicate any clear antiferromagnetic transition down to $T$=2K. Data collected in 1T are shown in Figure 3. The negative intercept of the inverse susceptibility with the temperature axis is found to decrease in magnitude with increasing measuring field. This behaviour indicates that the field is coupling to a ferromagnetic component in the Hamiltonian of Kapellasite. This may be due to either a field-dependent triplet state or the antisymmetric exchange that is allowed in the kagome antiferromagnet. In all fields, the temperature-dependence of the effective moment displays a marked reduction upon cooling below $T$~25K associated with a gradual moment collapse. Such behavior is not observed in the polymorph Herbertsmithite, and suggests that at low temperature the magnetic spins of Kapellasite condense into a non-magnetic singlet state. Further data, ideally collected with a local probe such as NMR or MuSR, is required to confirm whether the low temperature ground state of this KAFM is indeed the non-magnetic quantum state of the RVB model.

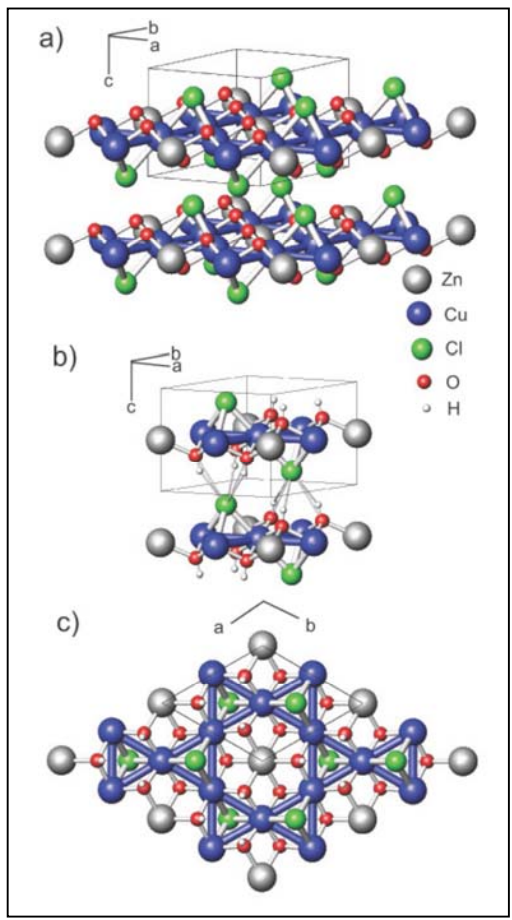

Figure 1. The quasi-2-dimensional lamellar structure of Kapellasite: a) without and b) with the weak interplane O—H—Cl hydrogen bonds; d) shows the structure of Kapellasite projected down the [001] axis and the kagome lattice made from segregation of the $Cu^{2+}$ and $Zn^{2+}$ ions. The nearest neighbor Cu—Cu distances in the kagome plane and between kagome layers in Kapellasite are 3.15Å and 5.73 Å, respectively..

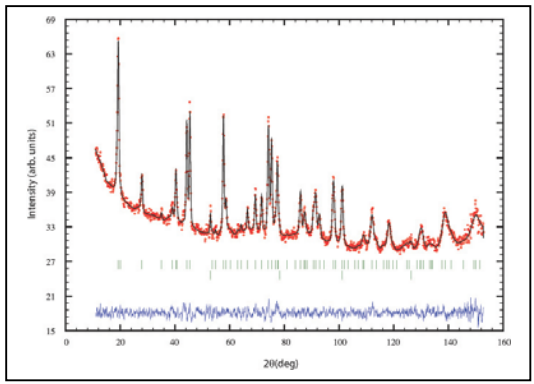

Figure 2. Fit to the powder neutron diffraction spectrum of Kapellasite, $Cu_3Zn(OH)_6Cl_2$, collected at room temperature over a period of 22 hours using neutrons of wavelength 1.909Å . The tick marks indicate the predicted locations of the nuclear scattering. The crosses correspond to the observed scattering, the line to the calculated diffraction pattern of the nuclear phase and the difference is shown below. The final goodness of fit parameter is $\chi^2=0.99$

Figure 3. Dc magnetic susceptibility data for Kapellasite collected with a SQUID magnetometer. a) The zero-field cooled (○) and field cooled (●) susceptibility taken in a measuring field of 10 000 G. b) The inverse susceptibility measured in a variety of fields: in low magnetic fields a strong deviation from the Curie-Weiss law is seen above 250 K; this deviation lessens when the measuring field is increased. The straight lines are guides to the eye for Curie-Weiss-type behaviour in 100, 500 and 70 000G. c) The effective moment, $\mu_{eff}=\sqrt{(8\chi T)}$, S=1/2 moments beginning to condense into a non-magnetic singlet state

In conclusion, we present the crystal structure and preliminary magnetic susceptibility data from the new $S=1/2$ kagome antiferromagnet Kapellasite, $Cu_3Zn(OH)_6Cl_2$. The kagome layers are only weakly coupled by O—H—Cl hydrogen bonds leading to highly 2D magnetic system. In contrast to its polymorph Herbertsmithite, Kapellasite displays a collapse in the effective moment on cooling below $T\sim25K$ suggesting that the spins of this quantum frustrated magnet are condensing into a singlet ground state.

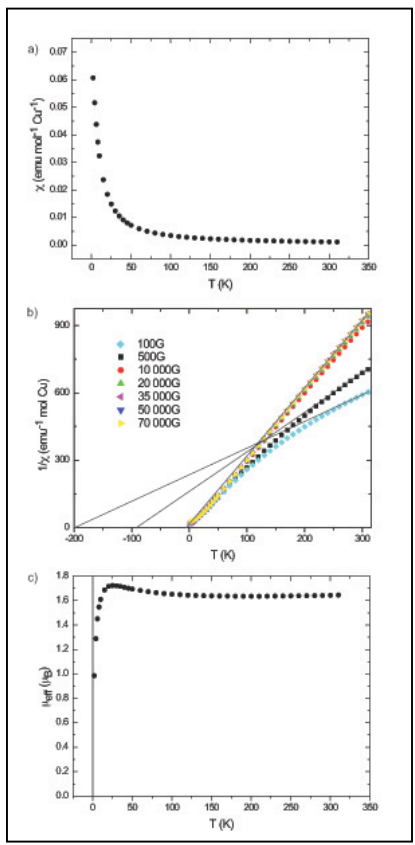

We thank the Royal Society and EPSRC (grant number EP/C534654) for financial support, and the ILL for provision of neutron time.

Supporting Information Available